\documentclass{article}
\usepackage{emulateapj,onecolfloat}

\input epsf
\lefthead{Levine {\em et al.}}
\righthead{Future Galaxy Cluster Surveys}

\begin{document}
\twocolumn[
\title {Future Galaxy Cluster Surveys: The Effect of Theory
Uncertainty on Constraining Cosmological Parameters}
\author {E.S. Levine${}^1$, A.E. Schulz${}^2$, Martin White${}^1$}
\affil{$^{1}$Department of Astronomy, University of California,
Berkeley, CA 94720}
\affil{$^{2}$Department of Physics, Harvard University,
Cambridge, MA 02138}

\begin{abstract}
\noindent 
Using the Fisher matrix
formalism, we quantitatively investigate the constraints on a 10
dimensional space of cosmological parameters which may be obtained
with future cluster surveys.  We explore the dependence of the
$\Omega_{\rm m}$ constraint on both angular coverage and depth of
field.  We show that in each case there is a natural cutoff beyond
which the constraints on $\Omega_{\rm m}$ do not significantly
improve.  We also investigate the sensitivity of the constraints to
changes in our knowledge of the Mass-Temperature (M-T) relation by
including its normalization and scatter as two of the parameters in
the Fisher matrix.  To make our analysis more realistic, we have
added, as priors, the Fisher matrices from hypothetical supernova and
CMB experiments.  We find that X-ray cluster surveys actually help to
constrain the M-T relation, and explore the implications of this result.

\end{abstract}

 ]

\rightskip=0pt
\section{Introduction} \label{sec:intro}

One of the most puzzling mysteries in cosmology today is the nature
of the dark energy, believed to be driving the accelerated expansion
of the universe.  This constituent is predicted to be smooth, except
possibly on horizon scales, so its cosmological effects are observed
solely through its influence on the expansion rate, $H$.  Since
the expansion is accelerating, the dark energy equation of state
$w_\phi=P_\phi/\rho_\phi<0$.  Assuming that the equation of state
varies slowly with time, the dark energy component must only recently
have become cosmologically significant. In order to best constrain
this component, it is therefore desirable to probe the
redshift range $0<z \lesssim2$.

Several authors have suggested using counts of clusters of galaxies
to probe the evolution of the dark energy in this redshift range
(\cite{hhm1}; \cite{AMI}; \cite{ref:HT2001}; \cite{ref:PR2001};
\cite{ref:NMCD2002}). Because these number counts depend on the growth function
as well as on the proper distance probed by Type Ia Supernovae (SNe) and Cosmic
Microwave Background (CMB) experiments, a cluster survey will yield a constraint
complementary to constraints from those experiments. Another advantage is
that clusters are big, bright, and sparse enough to make surveys of large
volumes relatively tractable.

Many factors affect our ability to extract constraints on
cosmological parameters from a cluster survey. In this paper we
examine the trade-offs in parameter estimation between conducting a
narrow deep survey which may require intimate knowledge of the nature
of the dark energy, and a broad shallow survey which may not. This is
because shallow surveys use the evolution of the cluster mass
function to constrain primarily $\Omega_{\rm m}$, and are much less
sensitive to $\Omega_{\phi}$ or the equation of state, $w_{\phi}$. In
this case, the interpretation of the cluster counts will be
relatively free of the modeling parameters. On the flip side, once
several of the other parameters are determined reasonably accurately
by combining a shallow survey with CMB and SNe constraints, a deep
cluster survey will be crucial if one hopes to measure any evolution
in the equation of state.

Considering the explosion of cluster survey activity that will occur
in the next few years, it will be important to either resolve or work
around the theory uncertainties in order to make the best use of the
imminent data.  One problem that must be examined is how to relate the
observable quantity (e.g.~cluster X-ray gas temperature) to the
quantity predicted by the cosmological theory (usually mass). In this
paper we determine quantitatively how uncertainties in the
Mass-Temperature (M-T) relation and in the nature of the dark energy
will affect the optimal constraints, and in light of that information,
identify the most effective types of cluster surveys to carry
out. Finally, accepting the current limitation of the theory, we show
how the constraints depend on the total number of clusters and
implicitly on the limiting mass of the survey. 

\section{Background} \label{sec:background}

There is a long history of using various cosmological tests to
constrain parameters in different directions of parameter space.
Often an observation will not be particularly useful in constraining
an individual parameter, but will rather nicely constrain a sum or
product of parameters.  The downside is that
usually there is another combination of parameters which is very
poorly constrained; a degeneracy orthogonal in parameter space to the
best constrained combination. To break the degeneracy, we can combine
the constraint with other cosmological experiments.  The key element
in obtaining a narrow region of likelihood contour intersection is
finding observables that each have a different dependence on the
cosmological parameters.

Much progress has been made in constraining the rather large set of
parameters typically present in today's cosmological models.  One of the
most import constraints comes from observations of the CMB.
Existing CMB experiments
(\cite{COBE}; \cite{TOCO}; \cite{BOOMERANG}; \cite{MAXIMA};
\cite{CBI}; \cite{DASI}) have already yielded information about the initial
conditions at the time of last scattering, and have made some progress in
determining the physical matter and baryon densities: $\Omega_{\rm m} h^2$,
$\Omega_{\rm b} h^2$, and the angular distance to the last-scattering
surface $r_\theta$.
Now with the upcoming {\sl MAP\/} and {\sl Planck\/} missions, the CMB
anisotropies will be studied with increasing precision, and the constraints
will be improved by incorporating new data on the polarization of the CMB.
The three primary parameters mentioned above will be measured to
accuracies approaching 1\%, with auxiliary parameters including the
equation of state $w_{\phi}$, measured to accuracies a few times worse
than this (\cite{HuEisTegWhi}).

The constraint on $r_\theta$ can be interpreted as essentially fixing
one combination of the matter and dark energy densities and the evolution
of the dark energy. However, even when we impose the condition
$\Omega_{\rm k}=0$, there are still degeneracies thats exist between
$\Omega_{\rm m}$ and the Hubble parameter, and to a smaller extent, the
equation of state of the dark energy, $w_{\phi}$.

There is a well known complementarity between the CMB constraints and
constraints from type Ia supernovae observations which one can take
advantage of to break these degeneracies (\cite{white1,snfish,PTM,HuEisTegWhi}).
The supernova constraints also come from a distance measure, but in
parameter space they are orthogonal to the constraints from the CMB
because the distance involved is much
smaller than the distance to the last scattering surface, from which all
of the CMB photons originate.  Thus for example, in measuring the energy
densities of the matter and the dark energy, the constraint from the Type
Ia Supernovae measures $\Omega_{\rm m} -\Omega_{\phi}$, whereas the CMB
measures a weighted sum of $\Omega_{\rm m} +\Omega_{\phi}$ (\cite{lambda}).
The effect is not quite as pronounced in the $\Omega_{\rm m} - w_{\phi}$
plane, but it seems clear that supernovae will provide a strong
complementary measurement.  The intersection of the likelihood contours
from a supernova experiment such as SNAP with CMB constraints obtained
using PLANCK could determine the matter density to 3.3\% if  the
polarization information is included (e.g.~\cite{snfish}).  An analysis 
of the optimal survey strategy for Supernovae searches has been carried
out by \cite{spergstark}, who determined that intermediate redshifts 
of $0.3-0.4$ are
the best ones to probe when combining the Supernova results with the CMB.

Although supernovae do seem to provide a clear way to narrow the parameter
space, it would be a very valuable cross check if a third observable was
found whose likelihood contours intersected the other two in the same
region of parameter space.  Such an independent constraint could additionally
remove one of the largest degeneracies in determining the equation of state
and its time evolution.
A promising observable to study is the number density  of rich galaxy
clusters as a function of redshift, an idea which was suggested as early
as 1992 by \cite{ref:OB1992} among others.
Studying the number density would provide an orthogonal constraint because
in addition to being sensitive to the comoving distance measure, it is
also sensitive to the rate of growth of fluctuations and hence the matter
density $\Omega_{\rm m}$ (in units of the critical density, $\rho_{\rm
crit}=3H_0^2/8\pi G$, where $H_0=100h\,{\rm km}\,{\rm s}^{-1}\,{\rm
Mpc}^{-1}$ is the Hubble constant). In models with a lower matter density,
the growth of structure ceases earlier. When normalized to the same level
of fluctuations today, this implies larger fluctuations at earlier times.
If the number density of objects is related to the amplitude of
fluctuations on a length scale $R\propto M^{1/3}$, as in the
Press-Schechter theory or its generalizations (\cite{PreSch}), then this
predicts more clusters of a given mass at high-$z$ in models with a low
$\Omega_{\rm m}$.  This method has been tested numerically by many authors
(\cite{ref:EFWD1988}, \cite{ref:ER1988}, \cite{ref:WEF1993},
\cite{ref:LC1994}, \cite{ref:GB1994}, \cite{ref:BM1996},
\cite{ECF}, \cite{ref:ST1999}, and \cite{jenkins}), and it was found that the
formalism accurately predicts the abundance of rich clusters as a function of
redshift for a broad range of different cosmologies.

Unfortunately, there are still many factors that present difficulties when
attempting to use a cluster survey to constrain the cosmological
parameters. A recent study (\cite{vianlid}) concluded that when all of the
major sources of error both in theory and observation are accounted for,
an unambiguous determination of the matter density $\Omega_{\rm m}$ is not
yet possible.  One complication is that the model of the evolution of the
dark energy equation of state is not well determined by any other
observations. This makes it difficult to interpret the impact of
observed evolution of the number density with redshift on the the
parameter $\Omega_{\rm m}$, because it becomes entangled with
$w_{\phi}(z)$.

Upcoming studies will introduce a golden age in cluster surveys. Several
observational programs will replace the current catalog of tens of
clusters at intermediate redshifts ($z<1$) with thousands of detections
improving not only the quantity, but the quality of the available data.
With the launch of the {\sl XMM/Newton\/} and {\sl Chandra\/} satellites,
the study of X-ray emission from the IGM has dramatically advanced.  The
XMM (X-ray Multi Mirror Satellite) Large-Scale Structure Survey
(\cite{XMMLSS})
will map the location of clusters and groups of galaxies out to $z\approx 1$
over an $8^\circ\times 8^\circ$ area.
This survey alone is expected to detect up to 800 clusters
in the temperature band ranging from 0.4 to 4 keV.
Another proposed survey using the same instrument, the XMM Cluster Survey
(XCS; \cite{Romer}), is expected to cover $\approx 800~\mathrm{deg}^2$ and
include more than 8000 clusters, as many as 1800 of which will have
well-defined temperatures
(estimates for $\Omega_{\rm m}=0.3, \Omega_\phi=0.7$) with $T>2$ keV.
There should even be a significant number detected with $z>1$.

Meanwhile, surveys for clusters are now being carried out at optical
wavelengths (using both galaxy counts and weak lensing) and using the
Sunyaev-Zel'dovich effect (\cite{SunZel}).
In the more distant future, the Planck satellite has the potential to
find $10^4$ clusters (with some dependence on cosmology) out to $z\sim 1$.
This enormous quantity of upcoming cluster survey work motivates our study
to find survey parameters that optimize the constraints on cosmological
parameters.

\vspace{0.5 cm}
\section{Method} \label{sec:method}

In order to address these questions we make use of the Fisher matrix
formalism (see \cite{fm} for a pedagogical review).
This allows us to elucidate the dependence of our hypothetical constraints
on both our survey design and our theoretical uncertainty, highlighting the
role of parameter degeneracies.
Because we have chosen to use the Fisher matrix, expressed in terms of the
derivatives of the likelihood function at the maximum, rather than a full
likelihood analysis, all of our constraints are ``local'' rather than global.
This should not affect any of our conclusions.

To make our analysis more relevant to the future where CMB and SNe surveys
have progressed significantly, we consider the constraints from a combination
of hypothetical observations.  First we have a cluster survey, but we include
as `prior' information the Fisher matrices obtained from a mock MAP mission
and a future SNe survey.  We describe each of these pieces below.

\subsection{Assembling the cluster Fisher matrix.}

We assume a flat cosmology and have chosen as our basis of 
parameters the space ($\Omega_{\rm m}=0.3$, $h=
.67$,
$\Omega_b h^2=0.018$, $n_S=1$, $w_{\phi}=-1$, $\delta_H=5.02\times 10^{-5}$,
$T/S=0$, $\tau=.2$,
M-T normalization $=6.59\times 10^{-10}$, $\sigma_{M-T}=.1$).
The quantities M-T normalization and $\sigma_{M-T}$ are defined below in
\S\ref{sect:temp}.

If we bin narrowly in the observable temperature and in redshift,
the abundance of clusters is naturally described by a Poisson distribution,
and so the Fisher matrix $\mathbf{F}$ becomes (\cite{hhm2})
\begin{equation} \label{eqn:ClusterFish}
{\mathbf{F}}_{ij}= -{ \partial^2 {\ln} {\mathcal{L}} \over \partial \theta_i
\partial\theta_j}
=\sum^{S}_{a=1}{\partial N_a\over \partial \theta_i}
                {\partial N_a\over \partial \theta_j}
                {1 \over N_a}
\end{equation}
where the sum runs over $S$ bins, chosen to be sufficiently small that our
fiducial model predicts $N_a\ll 1$ clusters per bin. In practice, binning in
redshift finer than $\Delta z=0.01$ does not change the constraints
significantly. Here $\cal{L}$ is the likelihood function, the $\theta_i$ are
the cosmological parameters and $N_a$ is the number of clusters expected in bin
$a$.
The derivatives are evaluated at the position of the fiducial model in
likelihood space logarithmically via the rule:
\begin{equation}
g'\left( \theta_i \right) = \frac{g(\theta_i \times \Delta\theta) - g(\theta_i
\div \Delta\theta)} {2\theta_i\ln(\Delta\theta)}.
\end{equation}
All derivatives were found to converge with $\Delta\theta=1.0001$.
When varying $\Omega_{\rm m}$ we also varied $\Omega_\Lambda$ to
maintain $\Omega_k =0$.
The dependence on the cosmological parameters enters through the mass
function $dn/dM$.  We use the fitting function of \cite{jenkins},
\begin{equation}
{dn\over dM} = - \frac{\rho_{\rm m}}{M^2} \frac{d \ln \sigma}{d \ln M}
\left[ 0.315
        \exp \left[-\left(-\ln \sigma + 0.61\right)^{3.8} \right]\right]
\end{equation}
where $\rho_{\rm m}$ is the mean background matter density, and $\sigma^2$ is
the variance of the mass, filtered on some scale $R(M)$.  It contains all the
details of the cosmological model through its dependence on the power spectrum
$P(k)$:
\begin{equation}
\sigma^2=\int_0^\infty 4\pi k^2\, dk\ \widehat{W}^2 (k) P(k)
\end{equation}

We have chosen a spherical top hat (in real space) as our smoothing window,
with Fourier transform $\widehat{W}$.  We use a power-law primordial power
spectrum and the fits to the transfer function of \cite{eandh}.

Using the M-T relation, along  with its associated Gaussian scatter, we map the
number densities by mass, $n(M)$, to number densities by temperature, $n(T)$.
We construct the Fisher matrix only from bins with a temperature greater than
a certain threshold; this  corresponds to the detection threshold of actual
surveys.  Though the majority  of our analysis is done with a threshold of
5 keV, we also explore the tradeoffs behind choosing different temperature
thresholds.
One benefit of this method  is that since the volume at a given redshift
scales linearly with angular survey  size, so too does the number of clusters
expected for a given temperature range.
Therefore, we have the relation $N_a = n_a V$, where V represents some number
of $(h^{-1} \mathrm{Mpc} )^3$ and $n_a$ is the number of clusters per
$(h^{-1}  \mathrm{Mpc} )^3$. This identity allows us to rewrite
(\ref{eqn:ClusterFish}):
\begin{equation}
{\mathbf{F}}_{ij} = V \sum^{S}_{a=1}
{\partial n_a\over\partial\theta_i} {\partial n_a\over\partial \theta_j}
{1\over n_a}.
\end{equation}
As $V$ is proportional to the angular survey area,
multiplying  the area by some factor corresponds to multiplying each Fisher
matrix element by  the same factor. We will take advantage of this below.

\subsection{Constructing the M-T relation} \label{sect:temp}

For the cluster $M-T$ relation we take the `virial' relation
(\cite{Kai86}; \cite{Pea})
\begin{equation}
T\propto M_{\rm vir}^{2/3} H^{2/3}(z) \Delta^{1/3}(z) h^{2/3}
\qquad .
\end{equation}
which can be derived by assuming clusters are self-similar, pressure
supported, isothermal spheres.
Here $M_{\rm vir}$ represents the virial mass of the cluster,
$\Delta(z)$ is the mean interior overdensity, relative to
the critical density, for which we say the object has virialized.
$\Delta$ is the familiar factor $18\pi^2$ for an $\Omega_{\rm m}=1$ universe,
and otherwise is given by the fitting function found by
\cite{BryNor}.

We fix the normalization of the M-T relation by appealing to numerical
simulations.
However, the normalization constant could potentially vary across different
cosmologies.  A comprehensive summary of all the different calculations of the
normalization is found in \cite{henry} and \cite{pierpaoli}.
Since we will compensate for differences in $h$, the parameters we
should be concerned with are $\Omega_{\rm m}$ and $\Omega_{\rm b}$.
We will use the value quoted by \cite{ENF} (hereafter ENF),
\begin{equation}\label{Treln}
T=6.59 \left(\frac{M}{10^{15} M_\odot}\right)^{2/3} \mathrm{keV}
\end{equation}
with a Gaussian scatter of $\sigma_{M-T}= 13.33\%$, because the
cosmology they simulate is very close to our fiducial model ($\Omega_{\rm m} =
0.3, \Omega_\Lambda =0.7, h=0.7, \Omega_{\rm b} = 0.04$).  Other than the
difference in $h$, we match these exactly. This normalization was calculated by ENF as an average
X-ray emission weighted temperature of 10 clusters at $z=0$.

Combining the normalization with the proportionality,
we arrive at the final form of our M-T relation:
\begin{eqnarray}
T=6.59\times 10^{-10}
\left(\frac{M_{\mathrm{vir}}}{M_\odot}\right)^{2/3}
\left(\frac{\Delta (z)}{\tilde{\Delta} (0)}\right)^{1/3} \nonumber \\
\left(\Omega_{\rm{m},0}(1+z)^3+\Omega_{\phi,0}(1+z)^{3(1+w)}\right)^{1/3}
\left(\frac{h}{0.7} \right)^{2/3} \label{eqn:mtreln},
\end{eqnarray}
where $\tilde{\Delta} (0)$ denotes the value of $\Delta(z)$ for the fiducial
model, and is not recalculated when we vary the cosmology to calculate
derivatives.  For now, we will assume that the evolution of the $M-T$ relation
is  described (perfectly) by Eq.~(\ref{eqn:mtreln}) over the range of  redshift
of relevance. We shall return to this point later.

\subsection{Assembling the CMB and SNe Fisher matrices}

To make our analysis more realistic, we have computed the Fisher
matrices associated with CMB and SNe experiments, and added them
as priors to the Fisher matrix obtained with a cluster survey.  To
compute the Fisher matrix associated with an experiment such as the
upcoming MAP mission, we have used the methods described in
\cite{ehandt} and have faithfully reproduced their constraints.
Assuming Gaussian perturbations and noise the
CMB Fisher matrix is given by
\begin{equation}
{\bf F}_{ij}=\sum_{\ell} \sum_{X,Y} {\partial C_{X {\ell}}\over
\partial \theta_i} \left( {\rm Cov}_{\ell} \right)^{-1}_{XY}
{\partial C_{Y {\ell}} \over \partial \theta_j}
\end{equation}
$C_{X \ell}$ is the power in the $\ell^{th}$ multipole and
the indices X and Y run over T,E,B,C; the temperature,
E-channel polarization, B-channel polarization and the
temperature- E mode cross correlation.
The matrix ${\rm Cov}_{\ell}$ depends on the $C_{X \ell}$, the
beam window function, and detector noise levels.  The explicit
form of the matrix can be found in \cite{ehandt} where we have
used the specifications for MAP at 90 GHz.  We have optimistically assumed
a sky coverage of 80\%.

Calculating the Fisher matrix for a SNe experiment is a problem
addressed recently in \cite{snfish}.  It depends on several
observables; the number of SNe observed, $N$, the mean
redshift of the observed SNe, $\bar z$, the scatter about this
mean redshift, $\Delta z$, and the standard deviation in the
observed magnitude from the predicted value $(\Delta m)^2$ .
Because we would
like to use the constraints that will be obtained from future experiments
such as SNAP, we also postulate a Gaussian distribution function $f(z)$ of
$N$ supernovae distributed about $\bar z$.
For a flat cosmology, the Fisher information matrix is given by
\begin{equation}
{\bf F}_{ij}={1 \over (\Delta m)^2} \int_0^{\infty} f(z)
w_i(z_n) w_j(z_n) dz,
\end{equation}
where
\begin{equation}
w_i(z)=\left( {5 \over {\rm ln} 10}\right) \eta^{-1} {\partial \eta \over
\partial \theta_i}.
\end{equation}
The $\theta_i$ are the cosmological parameters to be constrained, and
$\eta$ is given by
\begin{equation}
\eta(a;\Omega_{\rm m}, \Omega_{\phi})=\int_a^1 {da \over a^2 H(a)}
\end{equation}
where $ H(a)\equiv H_0 [\Omega_{\rm m} a^{-3} + \Omega_{\phi} 
a^{-3(1+w_{\phi})}]^{1/2}$.

We computed the SNe Fisher matrix for two different cases.
First, we calculated the constraints given by a cautious estimate of 200
supernovae. They followed a Gaussian distribution at a mean redshift of 0.65
with a dispersion of 0.3 in redshift. The standard deviation in the observed
magnitude was $\Delta m =0.3$. We also considered a more optimistic survey,
providing 400 supernovae with a mean redshift of 0.7 and a dispersion of 0.4
in redshift and $\Delta m=0.2$.  Both of these scenarios are relatively
conservative 
when compared to the expected yield of around 2000 Supernovae to be observed
by the proposed experiment SNAP. Both samples are used in the analysis below.

\section{Results} \label{sec:results}

It is instructive to compare Fisher matrices obtained using
different experimental setups.  In this paper, we change the
specifications of the cluster survey, and find the effect on
the constraints obtained.  An important question to consider is
whether the limitations on improving a constraint are dominated by the
experiment or the underlying theory, or some combination thereof.
In the analysis, we
have separated the two issues by parameterizing the
theoretical uncertainty in the M-T relation, and including its
normalization and scatter in
the cluster Fisher matrix.
To study experimental effects alone, we
simply invert the
Fisher matrix without any additional priors on the M-T relationship.
To examine the effect of theory uncertainties, we add additional priors
for the M-T normalization and scatter for a given
experimental setup.  Thus, we gain an intuition for which sources of
error are the most significant.

An observational question that merits attention when
designing the optimal cluster
survey is the trade
off between angular coverage on the sky, and depth of field of the
survey, $z_{\rm end}$.  It would be nice to know whether the bulk of the
constraint  on a parameter is determined by the redshift
evolution of the number density
of clusters, or rather simply on the total number of clusters today.
The former would suggest that a narrow, deep survey would be best, while
the latter could easily be accomplished with a shallow survey with a
wide angular coverage.  In general, the constraint depends on
both quantities, and the dominant effect depends upon the parameter
being considered.
We approach the question from the back by
simulating both types of observations and plotting the dependence
of the constraint on depth or angular coverage.

In our analysis of the various cluster surveys, we wish
to consider not only the survey's constraining capabilities on its own, but
also its ability to improve the constraints already obtained from
observations of the CMB and type Ia Supernovae.  
To obtain combined constraints, one has simply to
add the Fisher
matrices together and invert the resulting matrix.  In practice, using
Cholesky decomposition (\cite{press}) to calculate the inverse is somewhat
unstable because the fiducial parameters vary by many orders of magnitude.
This instability is avoided by performing the analysis on the logarithms of the
cosmological parameters.  Since the Fisher matrix transforms as a
covariant tensor
under a variable change in parameter space, this modification is easily
accomplished using the Jacobi matrix.

The $i$th diagonal element of the inverse of the sum of the three Fisher
matrices gives a lower bound on the $1-\sigma$ error bars obtained from 
all three experiments. When the other parameters are
allowed to vary:
\begin{equation}
\Delta \theta_{i}^2 \geq  ({\bf F}^{-1})_{ii} .
\end{equation}
This relation accommodates the propagation of errors in different
parameters,
since inverting the Fisher matrix is
equivalent to marginalizing over all of the other parameters.
We define the \textit{optimal percent constraint} as the minimum $\Delta
\theta_i / \theta_i$ allowed by equation 13, 
which is interpreted as the size of the
$1-\sigma$ error bar.

However, the Fisher matrix 
also quantifies the covariance of the parameters.  For each
type of observation, a 
realistic picture of its capabilities is obtained by plotting
elliptical curves of constant likelihood using a 2 parameter subspace of 
the inverted Fisher matrix. 
This view explicitly depicts the degeneracies discussed in
\S\ref{sec:background} between the two displayed parameters, while
marginalizing over all the rest.  
These degeneracies can be broken if the curves of constant likelihood
from other experiments are oriented in an orthogonal direction, since adding
the Fisher matrices effectively multiplies the likelihoods.
Figure 1 displays the $1-\sigma$ regions of likelihood for all
three individual experiments, and
also the combined constraint, illustrating the efficacy of this technique.
Table 1 contains the $1-\sigma$ errors on each of the parameters
after combining  the CMB matrix (using specifications of MAP with polarization),
200 supernovae  matrix
(about a mean redshift of 0.65), and a cluster survey of 1000 deg$^2$ with
temperature threshold 5 keV to a limiting redshift of $z_{\rm end}=0.7$.

Table 1 reveals an interesting phenomenon; the X-ray cluster
survey constrains  the normalization of the M-T relationship, if its redshift
evolution is assumed  known.
Without any prior information about the normalization of the the M-T curve from
simulations, a constraint of reasonable order can be achieved observationally.
The constraint on the scatter $\sigma_{M-T}$ is not as good.
This indicates that unless theory can predict the M-T normalization 
to better
than $\sim 20 \%$, prior knowledge of the normalization will not significantly
alter the results (see e.g. Pierpaoli, Scott, \& White 2001).
Prior knowledge of the scatter appears to be even less 
significant. We quantitatively investigate this below (see Figure 7).

\begin{figure}\label{allcons}
\centerline{\epsfysize=7cm \epsfbox{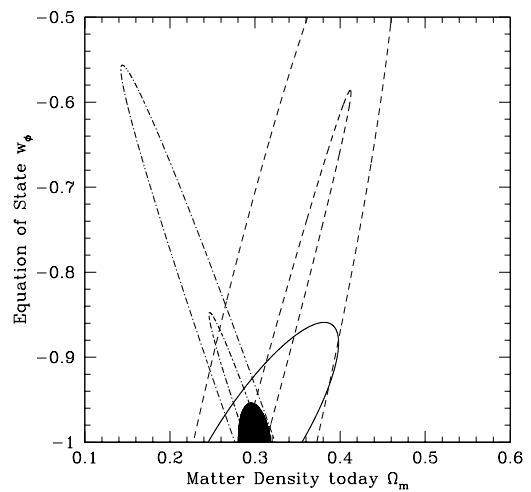}}
\caption{$1-\sigma$ regions of maximum likelihood in the $\Omega_{\rm m}-w$
plane.  Constraints come from a cluster survey of 1000 deg$^2$
to a final redshift of 1.2 and temperature threshold 5 keV (solid),
observations of the CMB (dashed)  using the specifications of MAP (outer) and
PLANCK (inner) including  the polarization information, and finally a
hypothetical supernova  observation (dot-dashed) with 200 (outer) and 400
(inner) supernovae.  The solid contour is the joint constraint from the cluster
survey, 200  supernovae, and MAP.}
\label{fig:joint}
\end{figure}

\begin{table}[t]
\begin{center}
\begin{tabular}{|r|r|r|}
Parameter &Fiducial value & $1-\sigma$ error \\ \hline
$\Omega_{\rm m}$ & 0.3 & 10.3\% \\
$h$ & 0.67 & 2.67\% \\
$\Omega_{\rm b} h^2$ & 0.0254 & 3.86\% \\
$n_S$ & 1 & 2.58\% \\
$w_\phi$ & -1 & 8.90\% \\
$\delta_H$ & $5.02 \times 10^{-5}$ & 5.05\% \\
$T/S$ & 0 &  $\pm$ 0.796\\
$\tau$ & 0.2 & 17.5\% \\
M-T norm & $6.59 \times 10^{-10}$ & 15.4\% \\
$\sigma_{M-T}$ & 0.1333 & 138\%
\end{tabular}
\end{center}
\caption[Table 1]{\rm The errors in each parameter from a combination of a CMB,
SNe, and cluster survey of 1000 deg$^2$ to a depth of z=0.7.
Percent errors are given in terms of the fiducial values.}
\end{table}

\begin{figure}
\centerline{\epsfysize=5cm \epsfbox{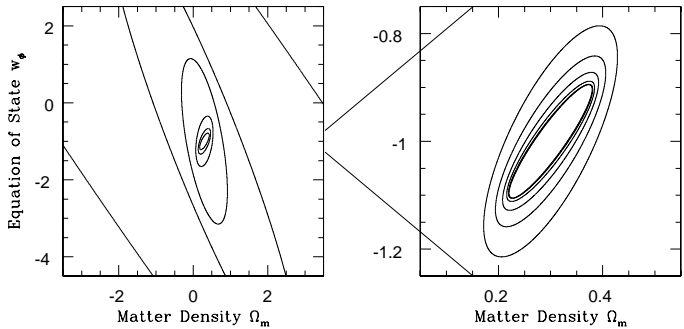}}
\caption{$1-\sigma$ regions of maximum likelihood in the
$\Omega_{\rm m}-w_{\phi}$ plane for a series of cluster surveys differing only
in survey depth.In the left hand panel, constraints are plotted for 
$z_{\rm end}=0.1, 0.2, 0.3, 0.5, 0.7, 0.9$ decreasing in size.  The right hand 
panel has $z_{\rm end}=0.9, 1.1, 1.3, 1.5, 1.7, 1.9$.}
\label{fig:ellipses}
\end{figure}

The orientation and size of the cluster ellipse in Figure \ref{fig:joint}
depends on the depth and angular extent of the survey. The change in orientation
in the $\Omega_{\rm m}-w_\phi$ plane has been noticed previously
(\cite{ref:Hu}). That the size will decrease with increasing depth is apparent;
more information will result in a better constraint. Figure \ref{fig:ellipses}
demonstrates how the orientation of the ellipse changes in a progression from a
shallow survey to a deep one at fixed angular size.  These plots do not contain
information from either the CMB or the SNe. However, the number density of
clusters at a given redshift is a quantity which is largely insensitive to the
parameter $\Omega_{\rm b} h^2$. This causes the Fisher Matrix to be almost
singular, reflecting the large degeneracy in that direction.
Thus we do rely on results from Big Bang Nucleosynthesis to provide us with
a 10\% prior on this quantity, which we include throughout our analysis for
greater numerical stability.
Note that the shape converges for the deepest surveys, as fewer and fewer
clusters above the detection threshold are detected.

The change in the orientation of the direction of degeneracy is a
direct consequence of the fact that cluster surveys probe a
combination of the proper distance and the growth function.  In the
range $0<z<0.5$, the proper distance to adjacent redshift bins is a
quickly varying function, and thus the sensitivity to small changes in
the cosmological parameters is dominated by changes in this proper
distance. Hence, the direction of degeneracy of these shallow cluster
surveys tend to align with contours of constant proper distance, as
can be seen in the first panel of Figure 2.  In the high redshift
range $z>1$ however, each redshift bin adds an increasingly large
number of clusters.  The number of collapsed objects at high redshift
is very sensitive to the growth function, and thus in this redshift
regime, the direction of degeneracy aligns instead with the contours
of constant growth factor, normalized to today.  This leads to an
interesting detail: populations of different masses will come to be
dominated by the growth factor at different redshifts.  This is
because of the hierarchical nature of structure formation.  More
massive clusters form later, and thus massive clusters get rarer
faster as deeper redshifts are probed.  The more massive populations get
dominated by the growth function at smaller redshifts.
We have determined that the higher the
temperature threshold, the lower the final redshift  $z_{\rm end}$
at which one runs completely
out of clusters; $z_{\rm end} \sim 1$ for 8 keV and $z_{\rm end} \sim 1.6$ for
5 keV (see Figure 2), 
while $z_{\rm end}$ is far beyond 2 if one stretches the threshold to
3 KeV.  The orientations of the constraints at these limits are also 
slightly different, being most orthogonal to CMB and supernova constraints 
for the smaller objects.  One complication is that the lower the temperature 
threshold is extended, the more one has to worry about extra physics that may 
affect the low mass systems, but are difficult to theoretically model.
These considerations must be traded off with depth and 
angular coverage when a survey is being planned.  The optimal 
strategies clearly depend on the capabilities of the surveying 
technique, which we discuss below. 

By far the easiest experimental specification to examine is the angular coverage
of the survey.  Changing the angular coverage effectively
changes the volume element at each redshift $z$.  As the Fisher
matrix contribution from the cluster survey is linearly
dependent on its volume, and the volume is, in turn, linearly dependent on
the survey's angular size, the parameter constraints solely from the
cluster survey go as:
\begin{equation} \label{eq:ErrorProp}
\Delta \theta \propto \frac{1}{\sqrt{S}}
\end{equation}
where $S$ is the solid angle subtended by the survey, where for a full sky
survey $S$ is $4 \pi / 41000$.
The volume element also depends on the comoving distance, which is
a function of the cosmological parameters. Figure 3 shows the improvement
in the constraint on the matter density $\Omega_{\rm m}$ for the three
experiments  as the angular size of the survey increases.  The relation is 
not quite as simple as in equation 14 because of the prior information provided
by the CMB and Supernovae.

\begin{figure}
\centerline{\epsfysize=7cm \epsfbox{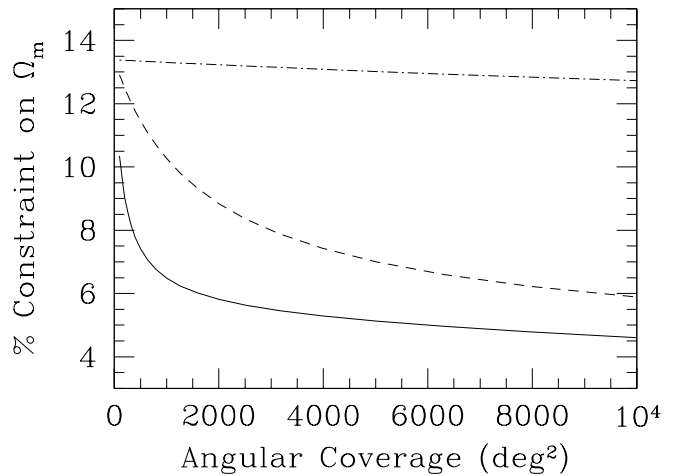}}
\caption{The percent constraint on $\Omega_{\rm m}$ as a function of the
angular coverage of the cluster survey.  We have added the fisher matrices
from the MAP CMB experiment, and from 200 supernovae, as well as
a 10\% prior on $\Omega_{\rm b} h^2$ from BBN.
The solid line is a survey to depth $z_{\rm end}=1.2$, the dashed
line is a survey to $z_{\rm end}=0.7$, and the dot-dashed line has a depth of
$z_{\rm end}=0.3$.}
\label{fig:Omvsarea}
\end{figure}

What we notice is that if the survey is deep enough
there seems to be a ``sweet spot'', a clear
cutoff place where the effort involved in performing the survey returns
maximum scientific information, and beyond which the payoff is not high.
For example, for survey depth $z_{\rm end}=1.2$ the payoff for extending the
observations  from 500 to 2000 deg$^2$ is quite considerable, but extending the
survey to  $10^4$ deg$^2$ does not improve the optimal constraint significantly
from a  survey of $\sim$ 2000 deg$^2$.  It is important to keep this turning
point  in the back of one's mind because increasing the angular coverage greatly
increases the total number of clusters that need to be observed.

Another strategy to consider is to make the survey
deeper rather than wider.  In this case there may be a smaller
total sample of clusters, but the evolution of the number density will
be more apparent.  In Figure 4, the dependence of the constraint on the
matter density is displayed as a function of $z_{\rm end}$, the depth
of the survey.  Of course many surveys are flux limited, and thus the 
depth of the survey is different for different mass scales; brighter 
more massive objects can be seen at greater distance.  However, since 
a significant portion of the constraint comes from the evolution of the 
density of clusters, and not just from the total number, it is crucial that
the sample be complete.  Therefore, the depth of the survey is effectively
the one at
which completeness is achieved for the faintest objects being counted, and
we will assume this in our analysis.  In practice this depth will depend
on the type of survey, and of course on the temperature threshold $T_{\rm min}$.
Constraints in Figure 4 are shown from the cluster survey alone, the
cluster survey combined with the expected constraint from the CMB MAP
mission (with polarization information), and finally from the cluster
survey combined with both the CMB and SNe experiments.

\begin{figure}
\centerline{\epsfysize=7cm \epsfbox{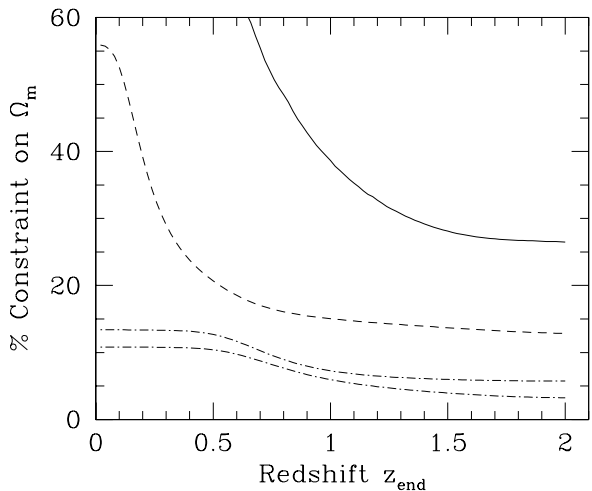}}
\caption{The percent constraint on $\Omega_{\rm m}$ as a
function of the depth of the cluster survey.
This survey has an angular coverage of 1000 deg$^2$.
The solid line is the constraint from the cluster survey combined with the
10\% BBN prior.  The  dashed line is obtained when the results of the CMB (MAP),
cluster observations, and BBN prior
are added, and finally the dot-dashed lines are the constraints
obtained when all three types of observations are used;
the cluster survey, the CMB, and either 200 (upper) or
400 (lower) SNe, also with the BBN prior.}
\label{fig:Omvsz}
\end{figure}

Again what we see is that there is a definite ``sweet spot.''  The role of
the prior information matrices of the SNe and CMB experiments is also
apparent in this graph.  The flat nature of both of the dot-dashed
curves between $z_{\rm end}=0$ and $z_{\rm end}\sim0.5$, which incorporate the
constraints  from all three types of experiments, are a result of the fact that
although  the survey is getting deeper and the
number of clusters detected is growing, the constraint from the cluster
survey is weak compared to combined
constraints from the CMB and SNe.  The solid
curve  displaying the cluster constraint alone shows that this is true.
In terms of the elliptical plots above, the
likelihood contours from CMB experiments would be entirely within
the region of maximum likelihood determined by the cluster survey.
After $z_{\rm end}=0.5$
we see a marked improvement, which means the likelihood contour will
have become small enough to intersect the CMB and SNe contour in
a way that decreases the most probable region compatible with all the
experiments.  Not only has the size of the cluster constraint decreased 
significantly by $z_{\rm end} \sim 0.5$, but as one can see from Figure 2
the orientation has passed the vertical and becomes increasingly orthogonal 
to the other constraints between $0.5<z_{\rm end}<1.7$.  
The role of the SNe constraint is also clear.  400
supernovae provide a narrower lever arm in parameter space than 200.
Since both SNe contours are at a significant angle to the cluster
constraints, and since they intersect at the position of the fiducial
model (by fiat), then the effect is
simply to decrease the area of the intersecting region by an approximately
constant amount,  which can be observed in Figure 4.

Another interesting property of experiments that should be carefully
studied is the effect of changing the detection threshold temperature on
the quality of the constraints that can be obtained.  In some types of
surveys this could be intimately related to the depth of the survey, for
example if the survey was flux limited, since flux correlates with temperature.
However this is not always true.  In the case of the SZE effect,
the limiting temperature is largely independent of redshift.
Also for X-ray surveys one could argue that those clusters for which accurate
temperatures can be known are well above the flux limit, allowing
essentially volume limited samples to be constructed.
Since temperature can be used as a proxy for mass, studying this dependence
is akin to asking whether most of the constraint comes from the rare high
mass clusters, or whether the large number of lower mass clusters are needed
in order to adequately constrain the parameters.
We would prefer the former scenario, because the higher $T_{\rm min}$, the
less sensitive are the results to non-gravitational physics in the modeling,
and the easier it is to make the observations.
The value of $T_{\rm min}$ has a strong nonlinear effect on the number of
objects that will be found within the domain of the survey, as Figure 5 shows.
Figure 6 illustrates the improvement in our knowledge of the matter density as
we lower the threshold temperature.

\begin{figure}
\centerline{\epsfysize=7cm \epsfbox{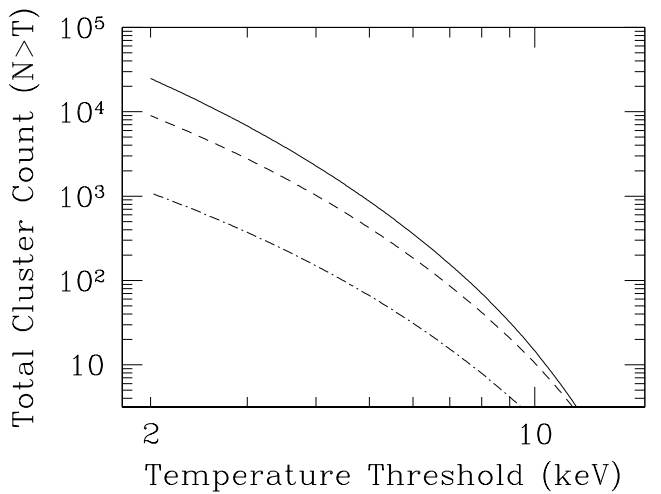}}
\caption{The total number of clusters detected at all redshifts out to
$z_{\rm end}=1.2$ (solid), $z_{\rm end}=0.7$ (dashed) and $z_{\rm end}=0.3$
(dot-dashed) as a function  of the detection threshold in keV for a fixed
angular aperture of  1000 deg$^2$.}
\label{fig:ngtt}
\end{figure}

\begin{figure}
\centerline{\epsfysize=7cm \epsfbox{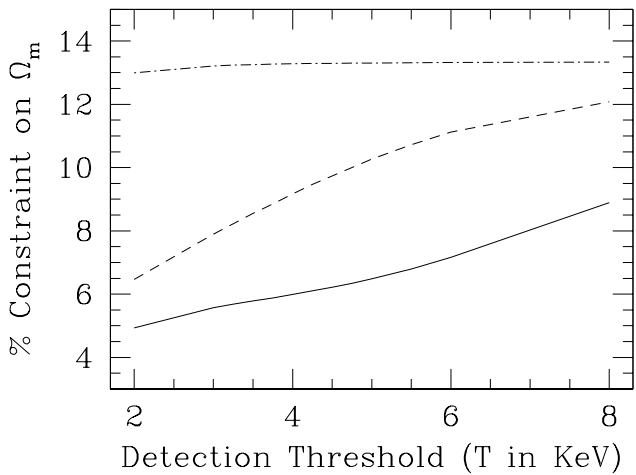}}
\caption{The percent constraint on $\Omega_{\rm m}$ as a function of the
temperature detection threshold.  We have added the same priors as in the
previous two figures.  The solid line is a survey to depth $z_{\rm end}=1.2$,
the dashed  line is a survey to $z_{\rm end}=0.7$, and the dot-dashed line has a
depth of $z_{\rm end}=0.3$.}
\label{fig:pomvstcut}
\end{figure}

This clearly demonstrates that it is indeed the rarest high mass clusters
that contribute the bulk of the constraint.  For a survey out to $z_{\rm
end}=1.2$, if  one drops the threshold halfway down the range of the plot, from
8 keV to 5 keV,  then one adds about 800 clusters and improves the
constraint by a factor of $3/4$ the original value. 
If one considers the bottom half from 5 keV to 2
keV,  one adds $\sim 24,000$ clusters, yet only improves the constraint 
by another factor of $3/4$. 
If  all clusters were equally important in constraining
$\Omega_{\rm m}$ then the constraint would drop significantly more
in the bottom half of the  plot, because we have increased the number of
clusters surveyed by more than an order of magnitude.

It is useful to probe how strongly the constraint on $\Omega_{\rm m}$ (or
any other parameter) depends on the theory implicit in the
survey.  Many cluster surveys involve observation of the X-ray luminosity
or temperature, either because they are X-ray surveys, or else they
are optical or SZE surveys that require an X-ray followup to obtain
redshift and temperature information.
For a cluster survey in X-ray temperature, this uncertainty is
framed in terms of the uncertainty in the normalization and the
scatter about the mass temperature relationship. It is conceivable that
adding more clusters to the survey will not affect the constraint on
$\Omega_{\rm m}$ simply because the Fisher matrix is dominated by the
error from the M-T normalization and scatter.  If this is the case,
considerable effort should be invested in improving these theoretical
considerations, so as to use the plethora of upcoming data to maximum
advantage.

By including the M-T normalization and scatter as parameters in the Fisher
matrix, we have investigated this important issue.  We find that
precise knowledge of the scatter about the M-T relationship does not
significantly improve the quality of the constraint on $\Omega_{\rm m}$.
Reducing the prior
on the scatter in a survey to $z_{\rm end}=0.7$ from a very loose prior of
100\%  to an  extremely tight 5\% improved the constraint on the matter density
less than  $1\%$, with a 10\% BBN prior and the CMB and 200 SNe
experiments included.  The improvement of the constraint on $\Omega_{\rm m}$
with better  knowledge of the M-T  normalization is of the same order when the
prior is changes from 10\% to 1\%. Weaker priors on the normalization have
little effect because the cluster survey constrains it to $~15\%$ by itself
(Table 1). Figure 7 shows this dependence for several redshifts.

\begin{figure} [!t]
\centerline{\epsfysize=7cm \epsfbox{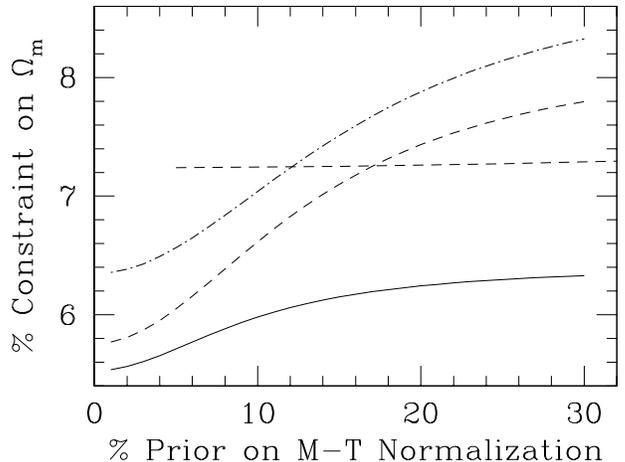}}
\caption{The percent constraint obtained for $\Omega_{\rm m}$ as a function of
the
prior knowledge of the error bar on the normalization 
of the M-T relation, also
reported in percent. The solid line is a survey to depth 
$z_{\rm end}=1.2$, the dashed
line is a survey to $z_{\rm end}=0.7$, and the dot-dashed line has 
a depth of
$z_{\rm end}=0.3$.  We have also shown the change of constraint as a function
of the prior knowledge of the scatter from the M-T relation for 
$z_{\rm end}=1.2$.  The constraint does not depend strongly on this quantity.}
\label{fig:normscat}
\end{figure}

\section{Conclusions} \label{sec:conclusions}

Many exciting surveys of galaxy clusters
have been proposed, involving several different observational techniques
to locate the clusters. In this
paper we have analyzed their potential to constrain two of the most fundamental
cosmological parameters, $\Omega_{\rm m}$ and $w_\phi$, within a larger
framework of  information gathered from other types of observations.
We have considered the effects of theoretical uncertainties on these constraints.
Even when the equation of state $w_\phi$ is postulated to be a constant,
large degeneracies exist in the $\Omega_{\rm m}-w_\phi$ subspace from all types
of observations
studied here.   We have varied
some of the characteristic attributes of cluster surveys such as the limiting
temperature
threshold, the depth of the survey, and the angular coverage,  and have noted
that the
direction of elongation of the elliptical region of maximum likelihood depends
considerably on these attributes.  Since we know the direction of degeneracy for
other important experiments, we can use this information to optimize the cluster
survey strategy, taking into account the various limitations for X-ray followup
observations.

We have deduced that in order to achieve maximal sensitivity to 
$\Omega_{\rm m}$, with minimal sensitivity to $w_\phi$
(without taking into account CMB and SNe
experiments), a fairly shallow survey to $0.4<z_{\rm end}<0.7$
(depending on the limiting temperature used) with large
angular coverage is an effective approach, because in this case the
constraint ellipse is vertically oriented in the $\Omega_{\rm m}-w_\phi$ plane.
It would be interesting for future work to see if such a constraint would be
relatively independent of $w_\phi'$. If so, such a measurement
would yield a significant lever arm to experiments such as SNAP, who would then
be in a good position to tighten constraints on $w'_\phi$. We also found that
for a survey of fixed angular extent,
the clusters with the highest constraining power are those with high temperature.
We confirm the claim made by \cite{hhm2} that it is the clusters past
$z>0.5$ that provide most of the constraint on $\Omega_{\rm m}$ (given our
assumption that $\Omega_{\rm m} + \Omega_\phi = 1$). This is especially
evident when SNe and CMB experiments are combined with the cluster survey,
because the part of the cluster constraint that is orthogonal to the SNe and
CMB ellipses comes from the clusters located at $z>0.5$.  We also see a
natural cutoff in limiting redshift beyond which the constraint on $\Omega_{\rm m}$
stops improving, as the number of clusters hotter than the detection threshold
falls off steeply with $z$.
More surprisingly, we also find that when the cluster survey is combined with 
the CMB and SNe constraints, increasing the field of view of the
survey does not improve the results indefinitely. 

In summary, when designing an experiment, several competing effects must be 
accounted for.  If one is interested in constraining $\Omega_{\rm m}$ and $w_\phi$
simultaneously, then
maximizing orthogonality with the CMB and Supernova constraints is desirable
because it will maximize the return for each unit of observing time that is spent
counting objects.  In general this suggests that deep surveys are advantageous
because it is at high redshifts that the growth function significantly affects 
the number density of clusters.
If an observable such as the SZ effect can truly probe out
to very high redshifts with relative completeness, then it is also advantageous
to stretch the temperature threshold as low as possible, since this allows
a deeper effective depth before the clusters run out. The resulting difference 
in angle is a relatively small
effect however, and naturally there are 
limits to the practicality of lowering the threshold. The number of dim distant 
objects that would need to be followed up grows exponentially as the threshold 
decreases.  To some extent, this can be compensated by making the survey narrower.

If, on the other hand, the survey is flux limited and can probe completely only
to a relatively shallow $z_{\rm end}$, such as an optical or X-ray survey of
clusters, the optimal strategy changes.  In this case we take advantage of the 
fact that very massive clusters are dominated by the growth function at smaller 
redshifts.  For maximum orthogonality in shallow surveys, one raises the temperature
threshold and studies only the most massive clusters.  We have shown 
that this will not significantly
weaken the survey's performance since typically the more massive
clusters contribute the bulk of the constraining power.  Of course to compensate
for the fact that huge cluster are rather rare, the survey can be widened, keeping
in mind that for each $z_{\rm end}$, there is a point of diminishing returns
when doing this.

In investigating different survey techniques' abilities to effectively whittle
down the likely region obtained from SNe and CMB experiments, we have been
conservative about the theoretical uncertainties involved in connecting the
observable (X-ray temperature) to the quantity predicted by the theory
(cluster mass).  We have included the normalization and scatter as parameters
in our Fisher Matrix analysis.
Any error in these parameters propagates to the constraints on the cosmological
parameters. 
We have investigated the sensitivity of our constraints on $\Omega_{\rm m}$
and $w_\phi$ to errors in the  normalization and scatter of the M-T relation
by including prior constraints on these quantities and observing the
improvement on the parameter constraints.  We found that in general, the
constraints on $\Omega_{\rm m}$ and $w_\phi$ are fairly robust against large
errors in the scatter of the M-T relation, but that they do depend somewhat more
strongly on errors in the normalization.
It is interesting that if no priors are added, the cluster survey itself will
constrain the normalization to around 10\%, suggesting that cluster surveys
could lead to some insight into intrinsic cluster properties.
We have not addressed what might happen if non-Gaussian effects were to
account for a large portion of the error on M-T, nor have we accommodated a
normalization that changes with redshift.
These would change the shape of the temperature-redshift distribution and could
impact the determination of the cosmological parameters profoundly.
To asses the impact of the first it would be necessary to employ a monte-carlo
method, rather than the simple Fisher matrix techniques used in this paper.
To understand quantitatively the question of evolution is beyond the scope of
this work.

It is clear that surveys of galaxy clusters will play a key role
in the future of further constraining the cosmological parameters.
Tight constraints on $\Omega_{\rm m}$ and the evolution of the dark energy
are essential to developing our models of the universe. A precise determination
of $\Omega_{\rm m}$ will help quantify the dark matter problem, and aid in
solving questions of structure formation.  The more $\Omega_{\rm m}$
can be independently
constrained, the better leverage we will have later when observations of
very deep sources begin to give us more information about the evolution of the
dark energy. 

\section*{Acknowledgments}

The authors would like to thank Daniel Eisenstein for his assistance
regarding the CMB Fisher matrix, and Saul Perlmutter for his useful insights.
Also thanks to J.D.Cohn, C. Metzler, and J. Mackey for the many clarifying
discussions on this work.
This research was supported in part by the Harvard College Research Program
(ESL), a Sloan Fellowship and the National Science Foundation.

\end{document}